\documentclass[12pt,a4paper,dvips]{article}

\usepackage{a4p}
\usepackage{cite,mcite}
\usepackage{graphicx}
\usepackage{epsfig}   
\usepackage{physics}
\usepackage{l3_title,ifthen}

 \date{September 15, 2003}

 \preprint{2003-060}

\newlength{\capindent}
\newlength{\capwidth}
\newlength{\figwidth}
\newcommand{\pho}{\phantom{0}}

\begin{document}

\begin{titlepage}

\title{Search for Doubly-Charged Higgs Bosons at LEP}

\author{The L3 Collaboration}

\begin{abstract}
  
Doubly-charged Higgs bosons are searched for in $\epem$ collision data
collected with the L3 detector at LEP at centre-of-mass energies up to
$209 \GeV$. Final states with four leptons are analysed to tag the
pair-production of doubly-charged Higgs bosons. No significant excess
is found and lower limits at 95\% confidence level on the doubly
charged Higgs boson mass are derived. They vary from $95.5 \GeV$ to
$100.2 \GeV$, depending on the decay mode.  Doubly-charged Higgs
bosons which couple to electrons would modify the cross section and
forward-backward asymmetry of the $\epem\ra\epem$ process.  The
measurements of these quantities do not deviate from the Standard
Model expectations and doubly-charged Higgs bosons with masses up to
the order of a Te\kern -0.1em V are excluded.

\end{abstract}

\submitted

\end{titlepage}

\section*{Introduction}

In the Standard Model of the electroweak
interactions~\cite{standard_model} the masses of the fermions and
bosons are explained by the Higgs mechanism~\cite{higgs_mech}. A
consequence of this mechanism is the existence of an additional
particle, the Higgs boson, which, to date, has not been directly
observed~\cite{l3higgs,lnq}. Extensions of the Standard Model predict
additional Higgs bosons which can be lighter and hence accessible at
current experimental facilities. Among these, doubly-charged Higgs
bosons, $\mathrm H^{\pm\pm}$, are
expected~\cite{Mohapatra:1981pm}
in several scenarios such as Higgs triplet models, left-right
symmetric models and, recently, little Higgs
models~\cite{littleHiggs}.

\begin{figure}[hp]
\begin{center}
\mbox{\epsfig{figure=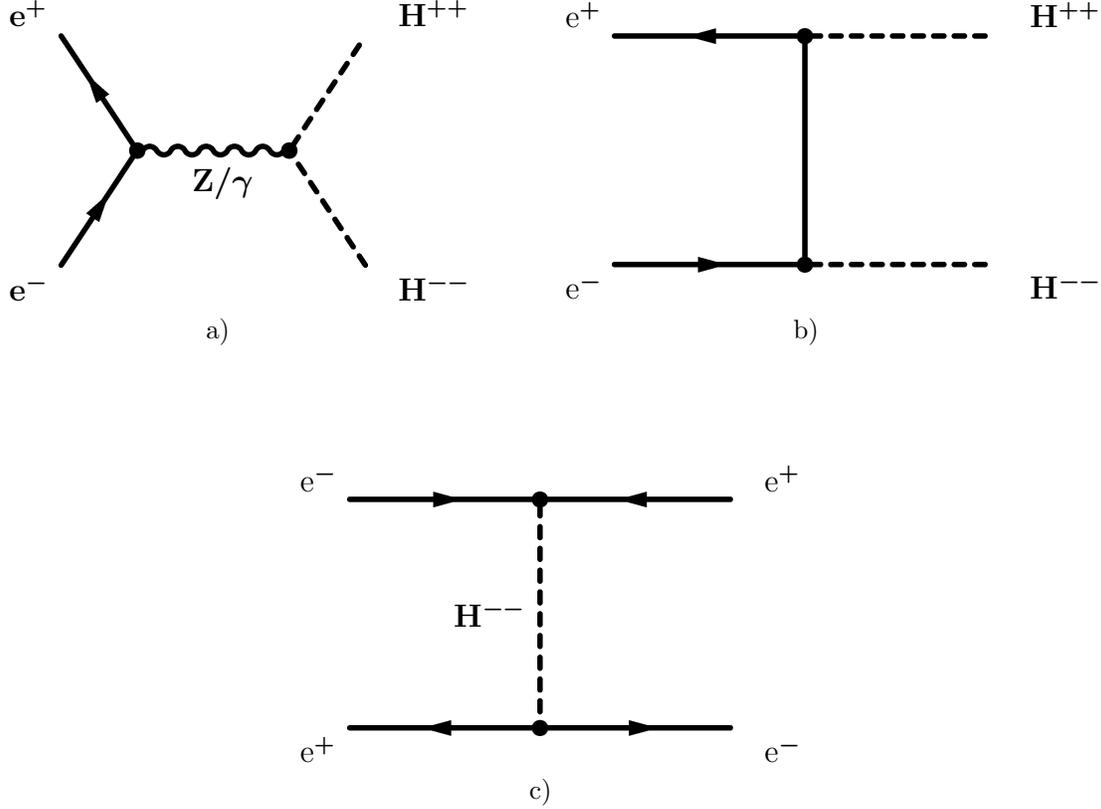,width=0.85\textwidth}}\\
\end{center}
\caption[]{\label{fig:1} a) $s$-channel and b) $t$-channel diagrams for
  the pair-production of doubly-charged Higgs bosons, c) 
   $u$-channel doubly-charged Higgs boson
exchange in the $\epem \ra \epem$ process.}
\end{figure}

Doubly-charged Higgs bosons can be light
enough~\cite{Aulakh:1998nn} to be
directly accessible in $\epem$ collisions at LEP through the
pair-production mechanism, depicted in Figures~\ref{fig:1}a
and~\ref{fig:1}b.  In addition, they can contribute to $\epem \ra
\epem$ scattering as sketched in Figure~\ref{fig:1}c, producing
measurable deviations in the cross section and forward-backward
asymmetries for masses of the order of a Te\kern -0.1em V.  This
Letter describes the direct search for pair-produced doubly-charged
Higgs bosons and the constraints derived from the precision
measurement of the $\epem \ra \epem$ scattering. Data collected with
the L3
detector~\cite{} at
centre-of-mass energies, $\sqrt{s}$, up to $209 \GeV$ are
used. Results from other LEP experiments were recently
reported~\cite{opal}.

The $\mathrm H^{\pm\pm}$ couplings to charged leptons are parametrised
by the parameters $h_{\ell\ell'}$, where $\ell$ and $\ell'$ denote the
charged lepton flavour. The search for pair-produced doubly-charged
Higgs bosons described below assumes $h_{\ell\ell'}>10^{-7}$ to
ensure that the $\mathrm H^{\pm\pm}$ decays before entering the
detector and $h_{\rm e\ell}<10^{-3}$ to suppress large contributions to
the cross section from the $t$-channel diagram of
Figure~\ref{fig:1}b. The latter assumption corresponds to a
conservative estimate of the experimental sensitivities.

Doubly-charged Higgs bosons are
conventionally labeled as ``left-handed'' or ``right-handed''~\cite{Mohapatra:1981pm}, referring to different couplings rather than different
helicities.  Left-handed $\mathrm
H^{\pm\pm}$ couple to the Z boson and the additional $s$-channel
diagram results in a pair-production cross section larger than for
right-handed $\mathrm H^{\pm\pm}$. The analysis discussed below
concentrates on the latter, less favourable, case.  The cross section
for the $\rm \epem\ra H^{++} H^{--}$ process
depends~\cite{Swartz:1989qz,Huitu:1998zt} only on the mass of the
doubly-charged Higgs boson, $m_{\rm H}$, and on $\sqrt{s}$. For
$\sqrt{s}=206 \GeV$, it varies from $\rm 1\, pb$ for $m_{\rm
H}=60\GeV$ down to $\rm 0.1\, pb$ for $m_{\rm H}=95\GeV$.

Pair-production of doubly-charged Higgs bosons produces events with
four charged leptons whose flavour depends on the $h_{\ell\ell'}$
coupling. In the following, all six possible couplings are considered:
$h_{\rm ee}$, $h_{\rm e\mu}$, $h_{\rm e\tau}$, $h_{\rm \mu\mu}$,
$h_{\rm \mu\tau}$ and $h_{\rm \tau\tau}$, with the hypothesis that
only one coupling at a time is different from zero, which implies
that both doubly-charged Higgs bosons in the events have the same
decay mode.

If the doubly-charged Higgs boson couples to electrons, it contributes
to the differential cross section of the $\epem \ra \epem$ process
through interference with the
additional $u$-channel Feynman diagram depicted in
Figure~\ref{fig:1}c. This additional term is
calculated~\cite{Swartz:1989qz} to be proportional to
\begin{displaymath}
{ h_{\rm ee}^2 \over m_{\rm H}^2 -u}
\end{displaymath}
where $u=- s(1+\cos{\theta})/2$ and $\theta$ is the electron
scattering angle. In the following, information on $h_{\rm ee}$ and
$m_{\rm H}$ is extracted from the comparison of the measured cross
section and the forward-backward asymmetry of the $\rm \epem \ra
\epem$ process with the Standard Model predictions and the doubly
charged Higgs contribution.

\section*{Data and Monte Carlo Samples}

The search for pair-produced $\mathrm H^{\pm\pm}$
uses $\rm 624.1\,pb^{-1}$ of data collected at $\sqrt{s}=189-209
\GeV$. Table~\ref{table:lumi} details the average $\sqrt{s}$ values
for the different data taking periods and the corresponding integrated
luminosities. Constraints on $\mathrm H^{\pm\pm}$ contributions to the
$\epem \ra \epem$ process are derived from these data and from an
additional $\rm 66.4\,pb^{-1}$ collected at $\sqrt{s}=130-183 \GeV$.

For the optimisation of the selection and efficiency studies, Monte
Carlo events of the process $\rm \epem \ra H^{++}H^{--} \ra
\ell^+\ell'^+\ell^-\ell'^-$ are generated according to the
differential cross sections of References~\citen{Swartz:1989qz}
and~\citen{Huitu:1998zt}.  Effects of initial state radiation are
included~\cite{Berends:1985dw} in the generation and final state
radiation is modelled with the PHOTOS~\cite{photos} Monte Carlo. For
each $\sqrt{s}$ value listed in Table~\ref{table:lumi}, several
$m_{\rm H}$ points are considered: $m_{\rm H}=45\GeV$ and from $m_{\rm
H}=65\GeV$ up to the kinematic limit $\sqrt{s}/2$, in steps of $5
\GeV$.  For each $m_{\rm H}$ point, 5000 events are generated for each
of the six $h_{\ell\ell'}$ couplings. Decays of the tau leptons are
described with the TAUOLA~\cite{tauola} Monte Carlo program and
JETSET~\cite{JETSET} is used to model hadrons produced in these
decays.

Standard Model processes are modelled with the following Monte Carlo
generators: KK2f~\cite{KK2f} for $\epem\ra\qqbar(\gamma)$,
$\epem\ra\mu^+\mu^-(\gamma)$ and $\epem\ra\tau^+\tau^-(\gamma)$,
BHWIDE~\cite{BHWIDE} for $\epem\ra\epem(\gamma)$,
EXCALIBUR~\cite{EXCALIBUR} for the four-fermion processes $\epem\rm\ra
q\bar{q}'e \nu_e$, $\epem\rm\ra \ell^+\ell^-q\bar{q}$ and $\epem\rm\ra
\ell^+\ell^-\ell'^+\ell'^-$, PYTHIA~\cite{JETSET} and
KORALW~\cite{KORALW} for four-fermion final states of the
$\epem\ra\Zo\Zo$ and $\epem\ra\Wp\Wm$ processes, respectively, which
are not covered by the EXCALIBUR simulations and PHOJET~\cite{PHOJET}
and DIAG36~\cite{DIAG36} for hadron and lepton production in
two-photon interactions, respectively.  The L3 detector response is
simulated using the GEANT program~\cite{my_geant} which takes into
account the effects of energy loss, multiple scattering and showering
in the detector.  Time-dependent detector inefficiencies, as monitored
during the data taking periods, are included in the simulations.

\section*{Search for Pair-Produced Doubly-Charged Higgs Bosons}

The signature of the $\rm \epem \ra H^{++}H^{--} \ra
\ell^+\ell'^+\ell^-\ell'^-$ process consists of four leptons, whose
flavour depends on the $h_{\ell\ell'}$ coupling. For electrons, muons
or leptonically decaying tau leptons this signature is clean and
little background is expected from lepton pair-production and
four-fermion processes. Events with tau leptons which decay into
hadrons have a larger background from the four-fermion $\epem\rm\ra
\ell^+\ell^-q\bar{q}$ process and from two-photon interactions. The
analysis proceeds from the identification of leptons to the
preselection of events compatible with the signal signature. Finally,
cuts on the lepton energies and global event variables further reduce
backgrounds.

Electrons are identified by requiring a well isolated cluster in the
electromagnetic calorimeter, formed by at least two adjacent crystals,
with an associated track in the tracking chamber. The shower shape of
this cluster must be compatible with that of an electromagnetic
particle.

Muons are reconstructed by requiring tracks in the muon spectrometer
matched with tracks in the central tracker. To reject cosmic
background, muon candidates must be in time with the beam crossing.

In addition to their leptonic decays, tau leptons are identified by
requiring low-multiplicity jets associated with one, two or three
tracks. Narrow and isolated jets are selected by comparing their
energy to that deposited in  $10^\circ$ and $30^\circ$ cones around
the jet axes.

To increase the selection efficiency, two additional classes of
particles are considered: photons, which correspond to electron
candidates which fail the track matching criteria, and minimum
ionising particles in the calorimeters, MIPs, having an associated
track in the central tracker, which tag muons.

Nine analyses are built which rely on the exclusive
identification of four leptons. They are denoted as: 
e\,e\,e\,e, 
e\,e\,e\,$\gamma$,
e\,e$\,\mu\,\mu$, 
e\,$\gamma\,\mu\,\mu$,
e\,e$\,\tau\,\tau$, 
$\mu\,\mu\,\mu\,\mu$, 
$\mu\,\mu\,\mu$\,MIP,
$\mu\,\mu\,\tau\,\tau$, and
$\tau\,\tau\,\tau\,\tau$. 
Each analysis is used in the study of one or more
$h_{\ell\ell'}$ couplings, as summarised in Table~\ref{table:analyses}.

In addition, three semi-inclusive selections are devised to increase
the selection efficiency for final states with tau leptons decaying
into hadrons. These selections first identify an electron or a muon
pair in hadronic events, including the case in which one of the
electrons is tagged as a photon, and then force the remaining
particles of the event into two jets by means of the
DURHAM~\cite{Durham} algorithm. These two jets are considered as tau
lepton candidates. The selections are denoted as: e\,e\,jet\,jet,
e\,$\gamma$\,jet\,jet and $\mu\,\mu$\,jet\,jet. They are used for the
analyses of the $h_{\rm e\tau}$, $h_{\mu\tau}$ and $h_{\tau\tau}$
couplings, as detailed in Table~\ref{table:analyses}.

\section*{Event Selection}

Low-multiplicity events with more than three but less than ten tracks
and visible energy in excess of $0.3\sqrt{s}$ are selected. Two
classes of events are accepted: events with at least three particles
identified as electrons, muons or tau leptons or events with two jets
and an electron or muon pair or one electron and one photon. The
numbers of events obtained by this preselection are given in
Table~\ref{table:selection}, where the results of the twelve different
analyses are combined and presented for the six $h_{\ell\ell'}$
couplings. Good agreement is observed between data and Standard Model
expectations.

\begin{figure}[hp]
\mbox{\epsfig{figure=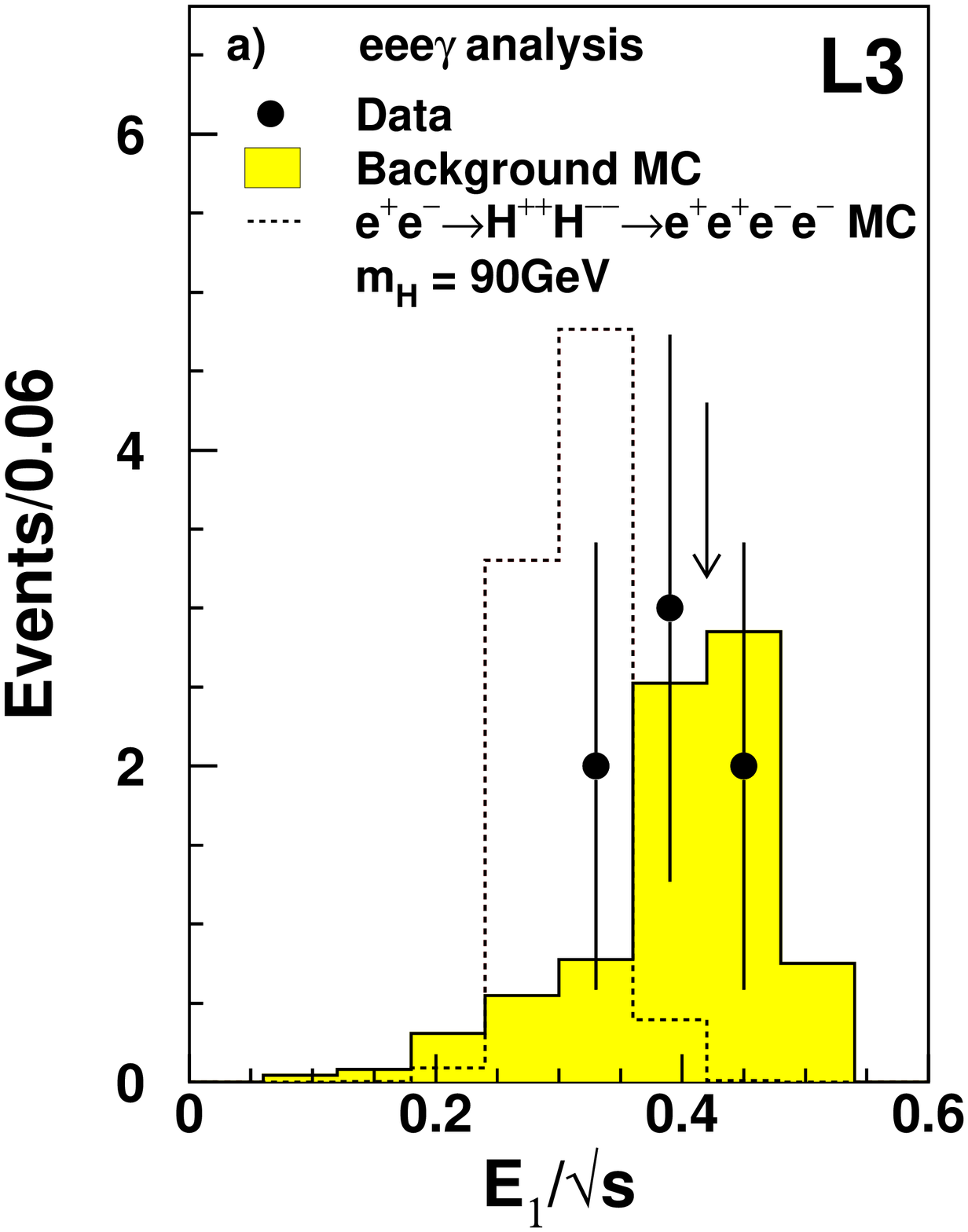,width=0.5\textwidth}%
      \epsfig{figure=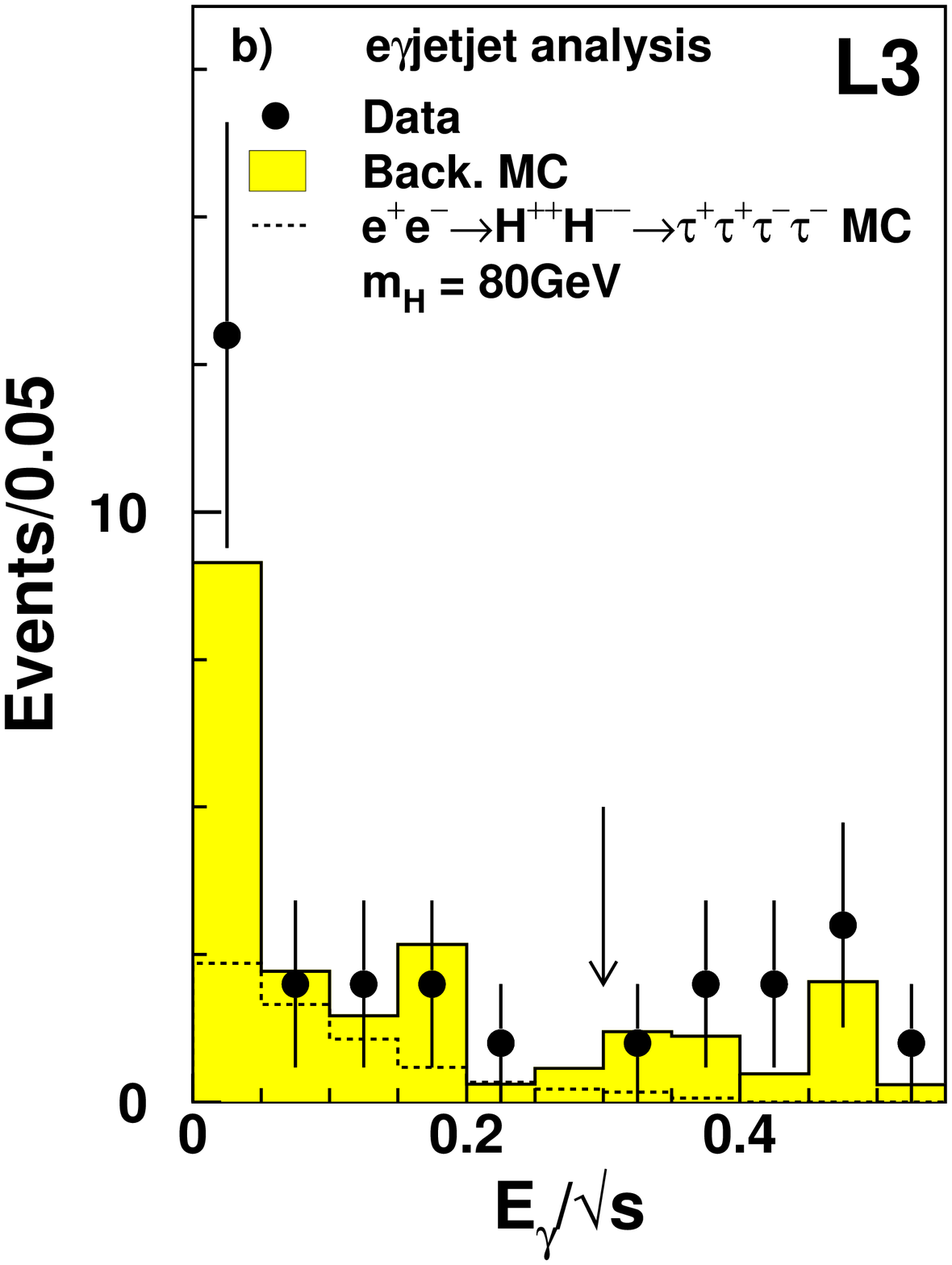,width=0.5\textwidth}}
\mbox{\epsfig{figure=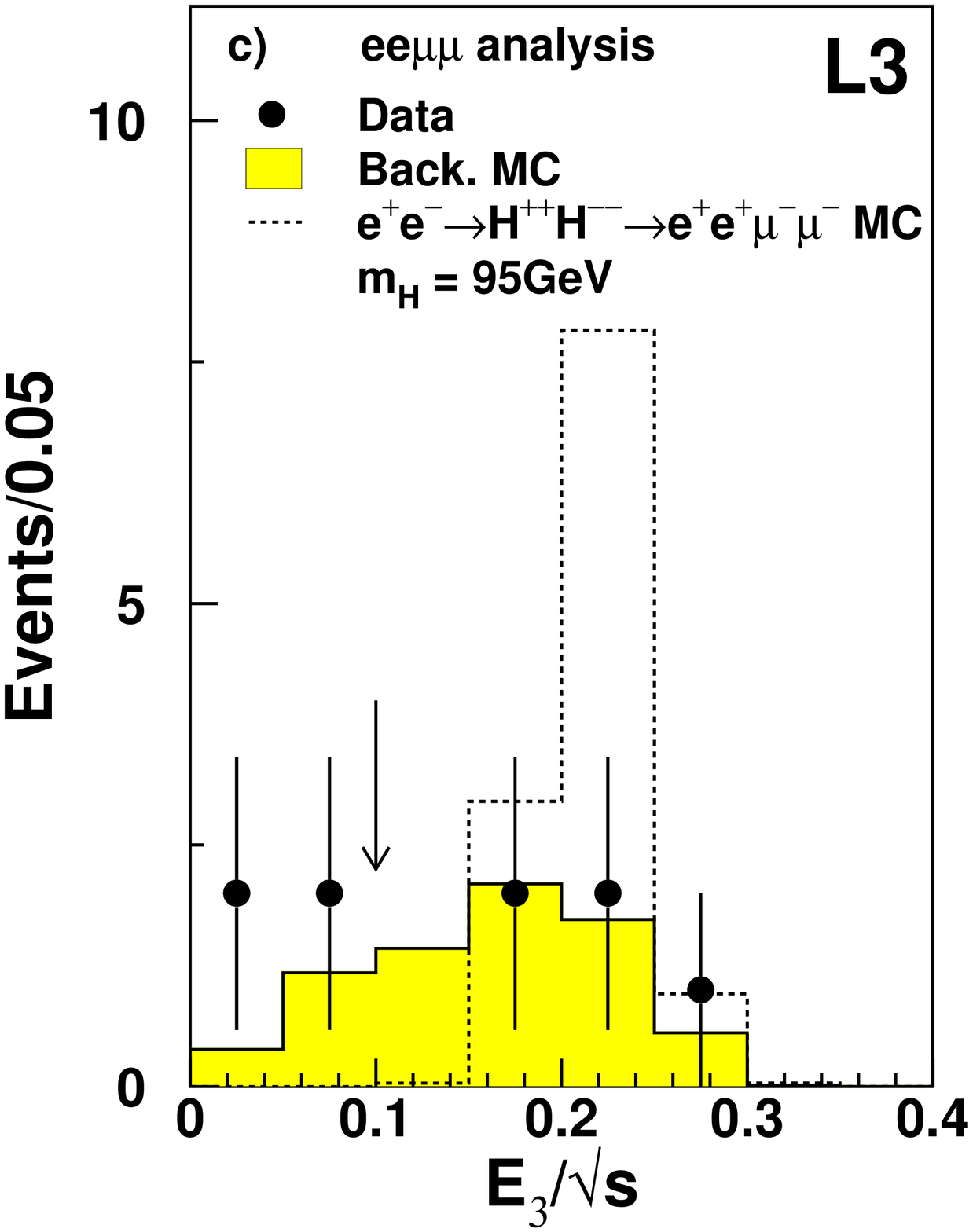,width=0.5\textwidth}%
      \epsfig{figure=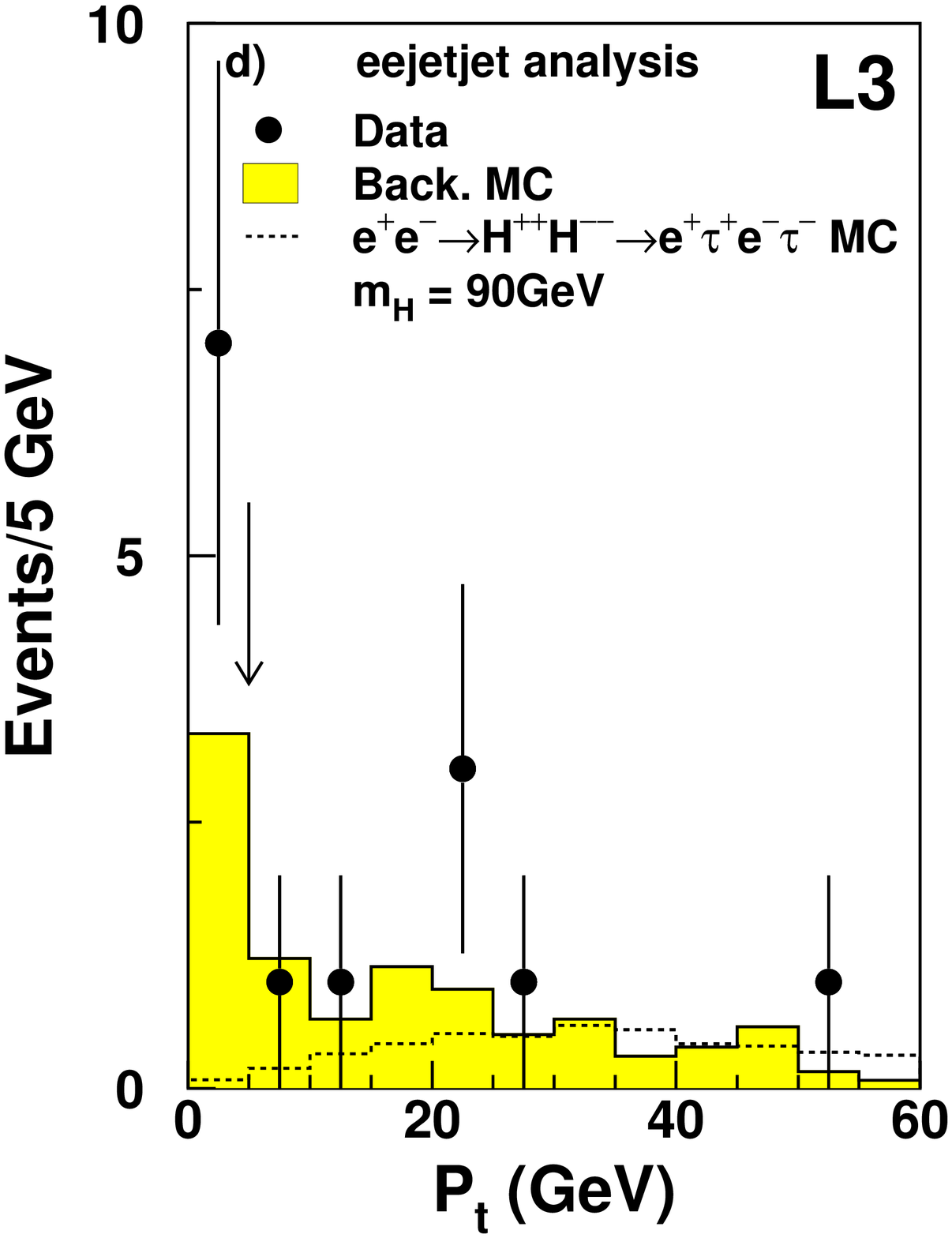,width=0.5\textwidth}}
\caption[]{\label{fig:2} Distributions for data, signal and background
  Monte Carlo of: a) the energy of the most energetic lepton in the
e\,e\,e\,$\gamma$ analysis, b) the photon energy for the 
 e\,$\gamma$\,jet\,jet analysis, c) the energy of the third most
  energetic lepton for  the e\,e$\,\mu\,\mu$
  analysis and d) the event transverse momentum for the
  e\,e\,jet\,jet analysis. The arrows indicate the position of the cuts.}
\end{figure}

Several discriminating variables are considered to increase the
sensitivity of the analysis.
\begin{itemize}

\item The energy of the most energetic lepton, $E_1$, is close to
  $0.5\sqrt{s}$ for the background from two-fermion events, and peaks
  around $0.25\sqrt{s}$ for the signal, which predicts a similar
  energy sharing for all leptons of the event. A  cut around
  $E_1<0.45\sqrt{s}$ is used by all twelve selections. As an example, the
  distributions for the e\,e\,e\,$\gamma$ analysis are shown in
  Figure~\ref{fig:2}a.

\item The energy of the second most energetic lepton  tends to
  be high for background events and peaked around $0.25\sqrt{s}$ for
  the signal. A cut around $0.35\sqrt{s}$ is applied for the
  e\,e\,jet\,jet and $\mu\,\mu$\,jet\,jet analyses.

\item The energy of the selected photon, $E_\gamma$, for initial state
  radiation photons from fermion pair-production has a high energy
  tail, as shown in Figure~\ref{fig:2}b for the e\,$\gamma$\,jet\,jet
  selection. A cut around $E_\gamma<0.3\sqrt{s}$ is applied for all
  analyses which accept photons. For events of the e\,e\,e\,$\gamma$
  and e\,$\gamma\,\mu\,\mu$ analyses, an additional cut around
  $E_\gamma>0.2\sqrt{s}$ is applied, to enforce the signal topology
  which predicts lepton energies around $0.25\sqrt{s}$.

\item The energy of the third most energetic lepton, $E_3$, is low
  for the background from two-fermion processes and
  non-resonant or single-resonant four-fermion production and also
  peaks around $0.25\sqrt{s}$ for the signal. A cut around
  $E_3>0.1\sqrt{s}$ is applied for the e\,e$\,\mu\,\mu$ selection, whose
  distributions are shown in Figure~\ref{fig:2}c. 

\item Events with jets in the final state suffer from a potentially large
  background from two-photon processes. This is reduced by requiring
  that an energy less than $30 \GeV$ is deposited in the calorimeters in a
  $30^\circ$ angle around the beam line and the projection of the
  missing momentum vector on this direction is less than $50 \GeV$.  The
  presence of neutrinos in tau lepton decays gives signal events a
  momentum imbalance in the plane transverse to the beam axis, $P_t$, as
  shown in Figure~\ref{fig:2}d for the e\,e\,jet\,jet analysis. A cut
  $P_t>5 \GeV$ is applied, further reducing events
  from fermion pair-production and two-photon processes which have
  small values of  $P_t$.

\end{itemize}

The twelve selections listed in Table~\ref{table:analyses} are
simultaneously applied and their yields are combined for the six
couplings. The nine selections without jets in the final state are
largely complementary, while a large overlap is observed between the
e\,e\,jet\,jet and e\,e$\,\tau\,\tau$ selections. Additional
selections like e\,e\,$\gamma\,\gamma$ and $\mu$\,MIP\,jet\,jet are
found not to increase the signal sensitivity.

\section*{Results and Interpretation}

Table~\ref{table:selection} compares the number of events observed
after final selection with the Standard Model expectations. Good
agreement is observed and no evidence is found for a signal due to
doubly-charged Higgs bosons. The number of expected signal events for
$m_{\rm H}=95\GeV$ and the selection efficiencies for the range
$m_{\rm H}=60-100\GeV$ are also given.

\begin{figure}[hp]
\begin{center}
\epsfig{figure=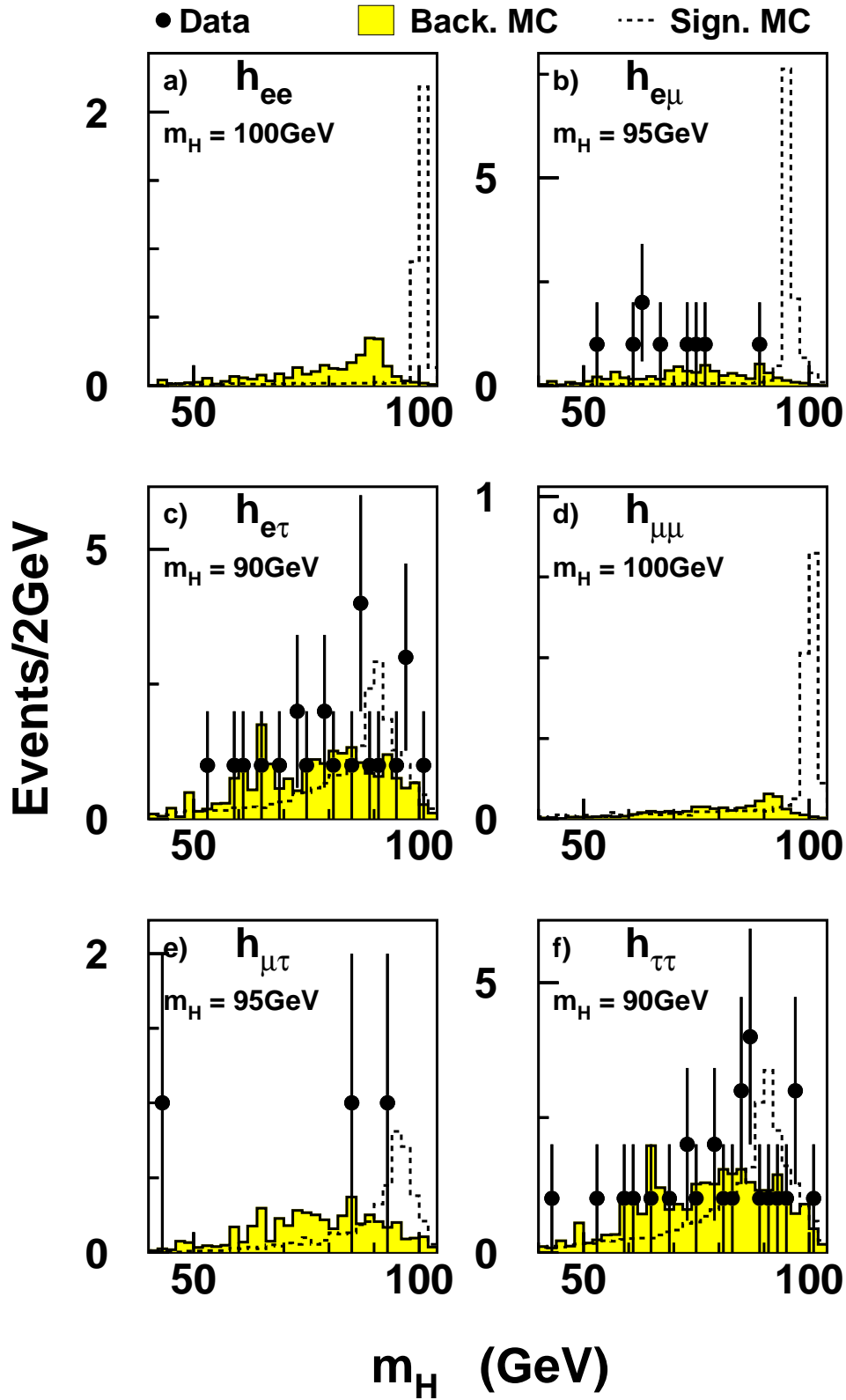,width=0.9\textwidth}
\end{center}
\caption[]{\label{fig:3} Distributions for data, signal and background
  Monte Carlo of the reconstructed Higgs mass for the: a) $h_{\rm ee}$,
  b) $h_{\rm e\mu}$, c) $h_{\rm e\tau}$,  d) $h_{\rm \mu\mu}$,
  e) $h_{\rm \mu\tau}$ and f) $h_{\rm \tau\tau}$ couplings.}
\end{figure}

The sensitivity of the analysis is enhanced by the reconstruction of
the mass of the candidate Higgs bosons. For each coupling, all
pairings of leptons with a flavour consistent with doubly-charged
Higgs boson decay are considered and their invariant and recoil masses
are calculated. The pairing with the smallest difference between these
two masses is retained and their average is used as an
estimate of $m_{\rm H}$. The distributions of the reconstructed mass
for data, Standard Model and signal Monte Carlo are presented in
Figure~\ref{fig:3}. No structure possibly due to a doubly-charged
Higgs signal is observed.

\begin{figure}[hp]
\begin{center}
\begin{tabular}{cc}
\mbox{\epsfig{figure=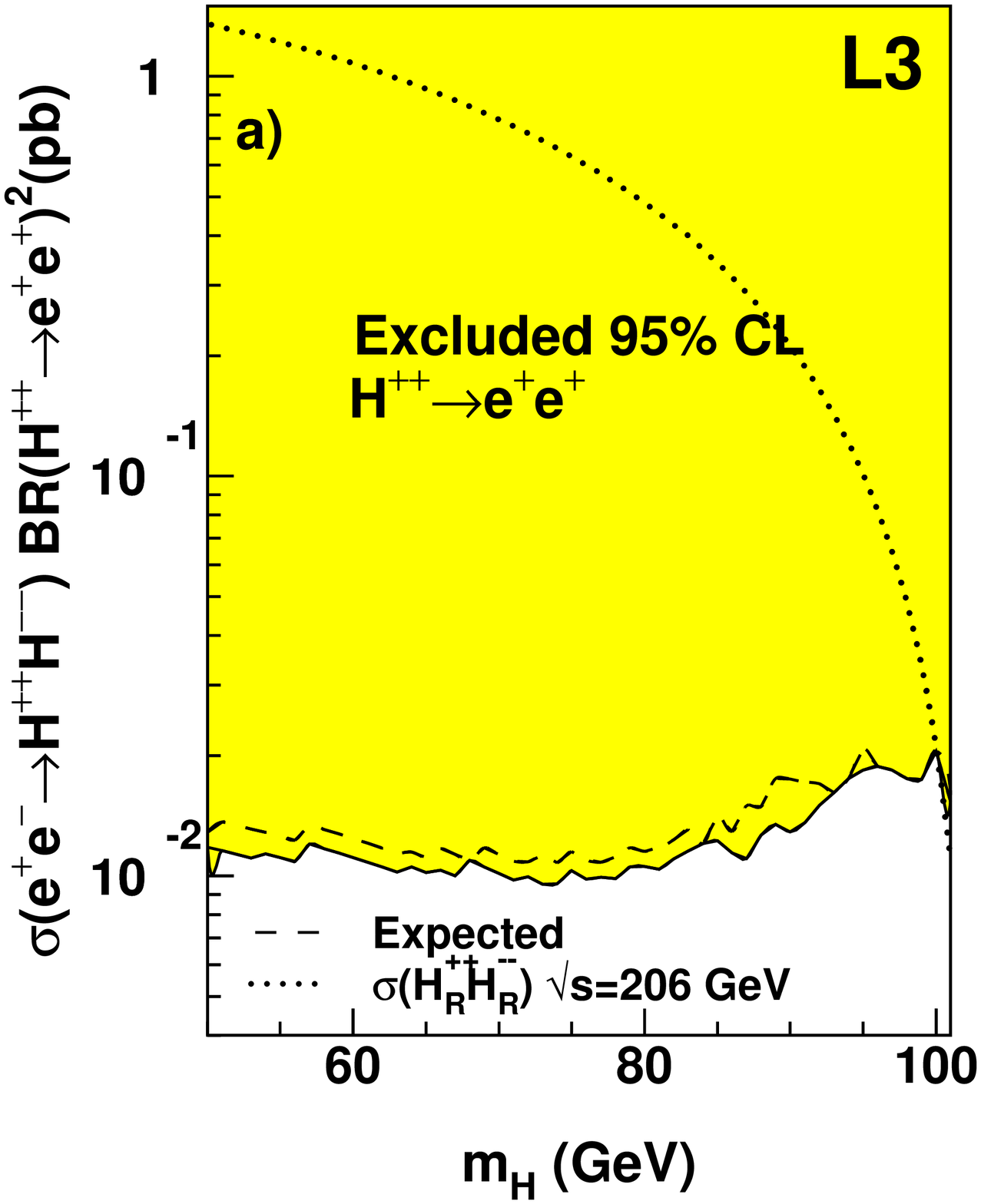,width=0.5\textwidth}}&
\mbox{\epsfig{figure=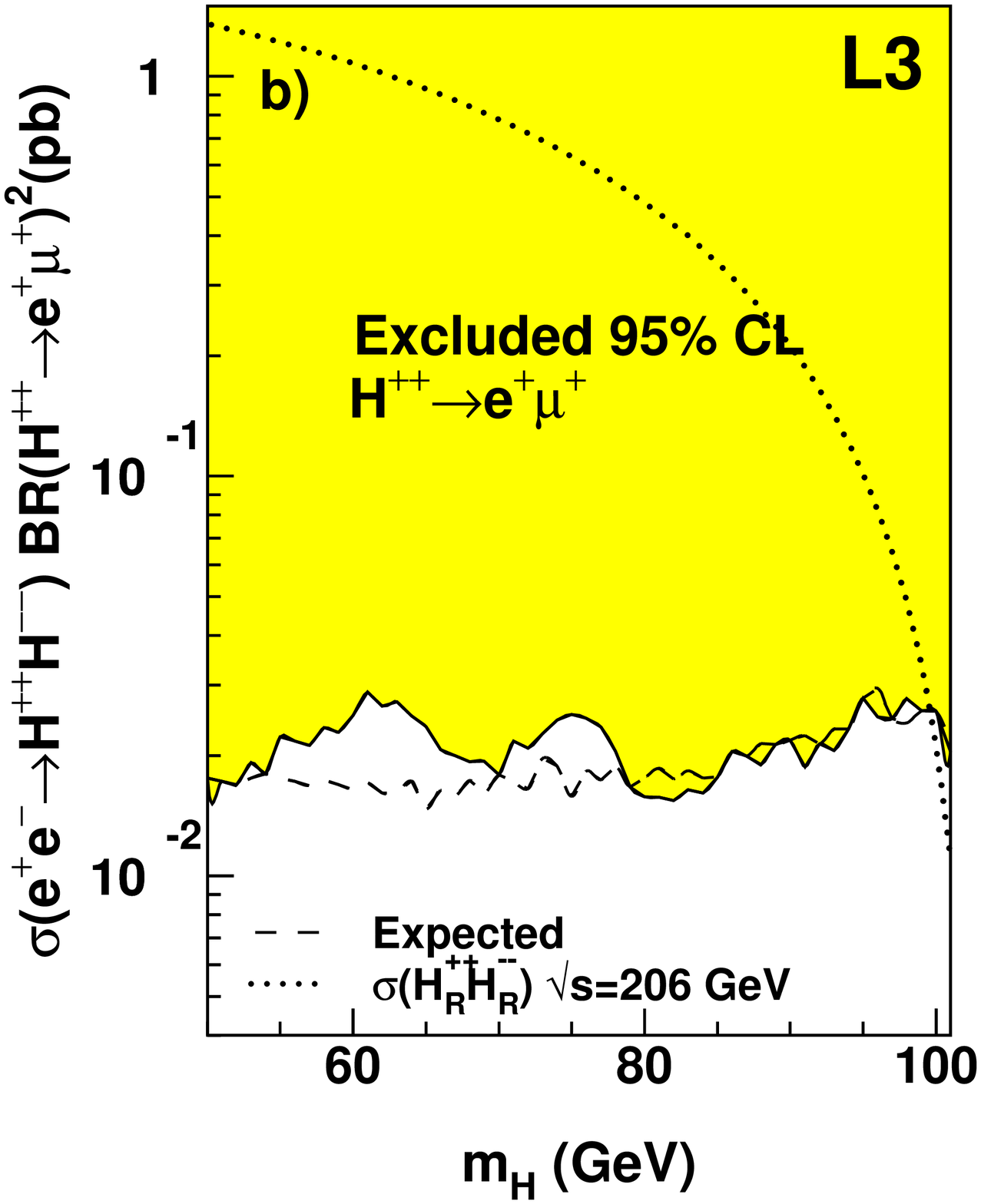,width=0.5\textwidth}}\\
\mbox{\epsfig{figure=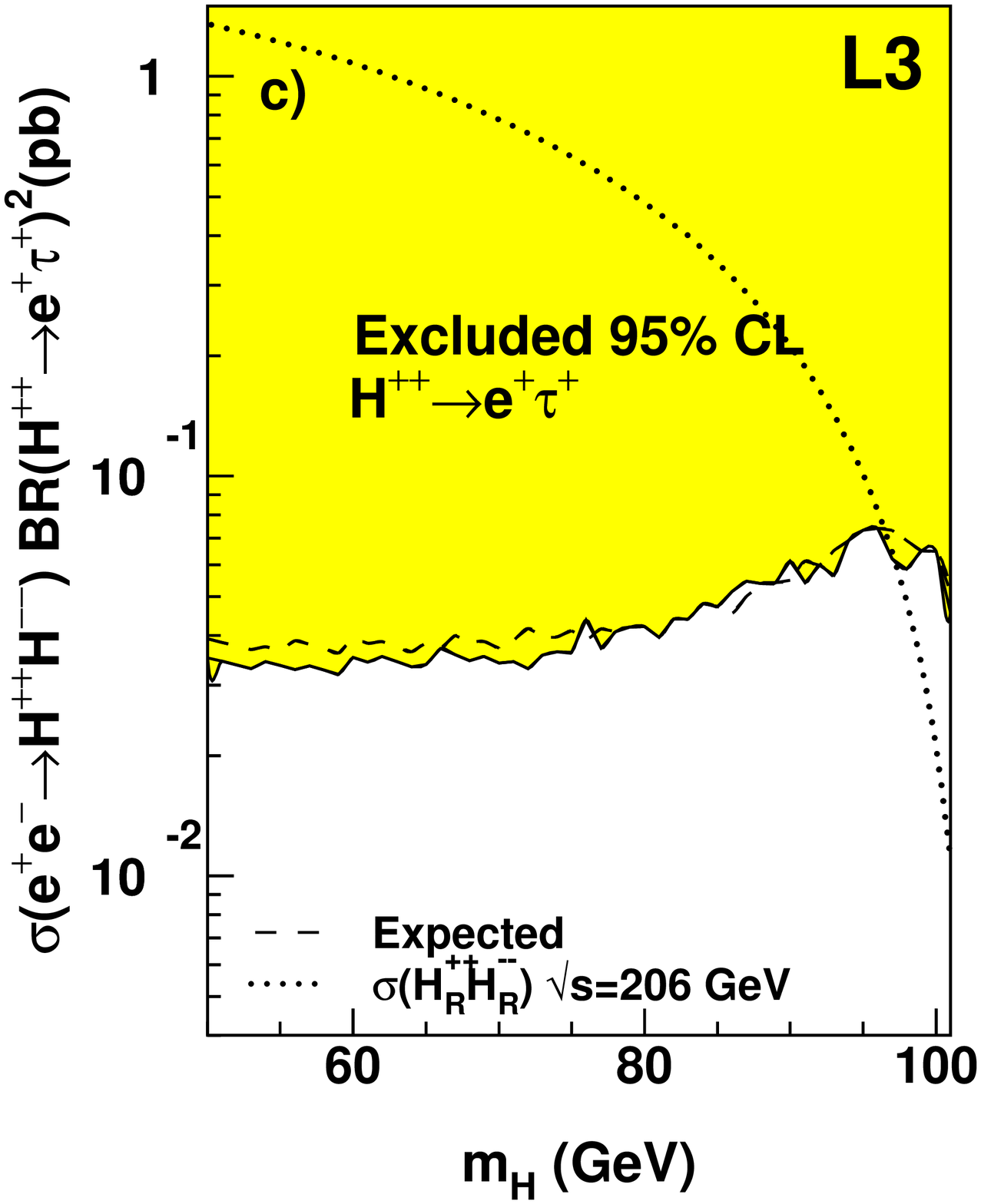,width=0.5\textwidth}}&
\mbox{\epsfig{figure=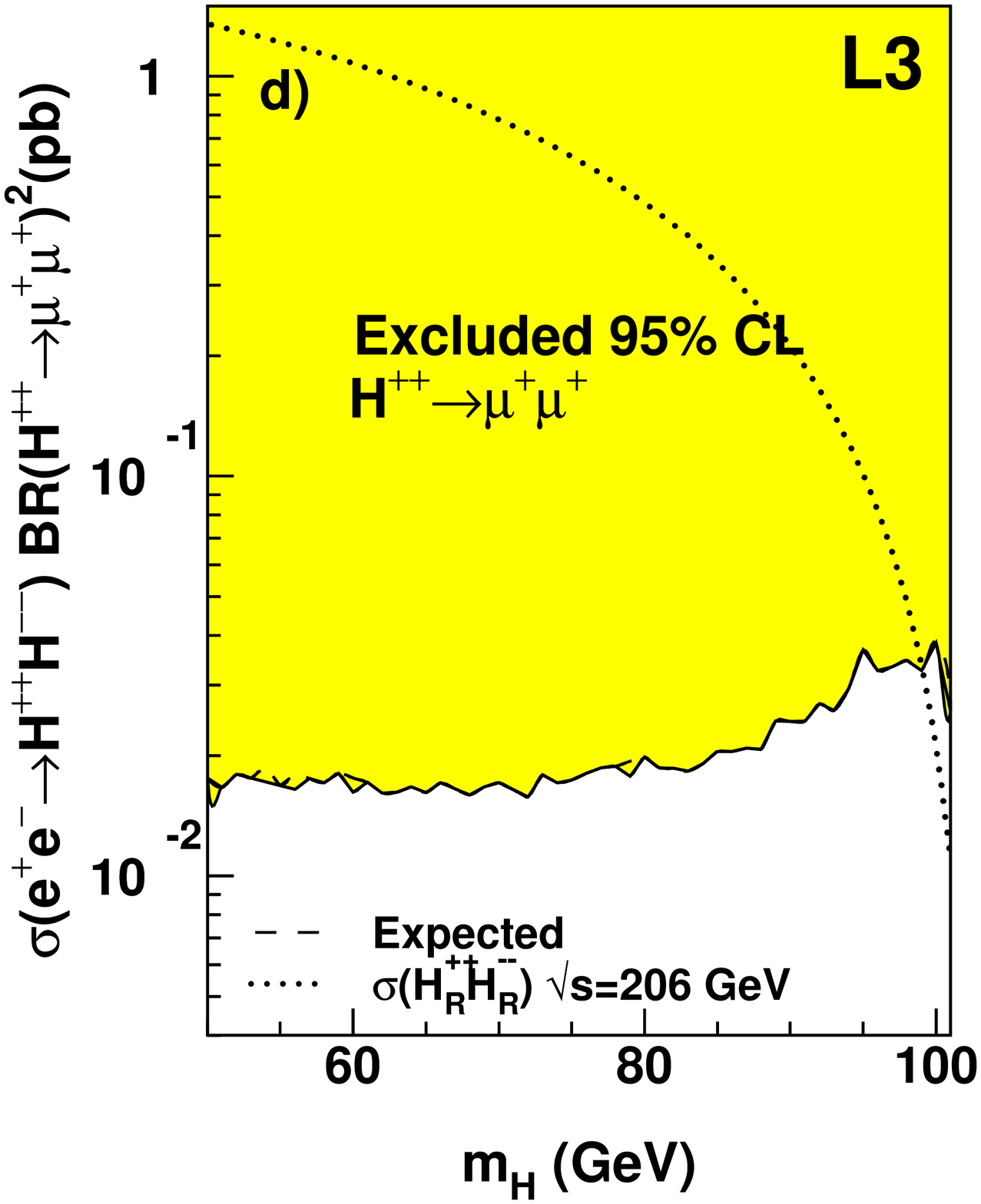,width=0.5\textwidth}}\\
\end{tabular}
\end{center}
\caption[]{\label{fig:4} Observed and expected limits on the cross
  section of doubly
  charged Higgs boson pair-production times its branching ratio in a given
  final state  as a function of $m_{\rm H}$ for the: a) $h_{\rm ee}$,
  b) $h_{\rm e\mu}$, c) $h_{\rm e\tau}$ and d)  $h_{\rm \mu\mu}$
  couplings. The expected cross section for the $s$-channel
  production of a right-handed doubly-charged Higgs boson is also shown.}
\end{figure}

\begin{figure}[hp]
\begin{center}
\begin{tabular}{cc}
\mbox{\epsfig{figure=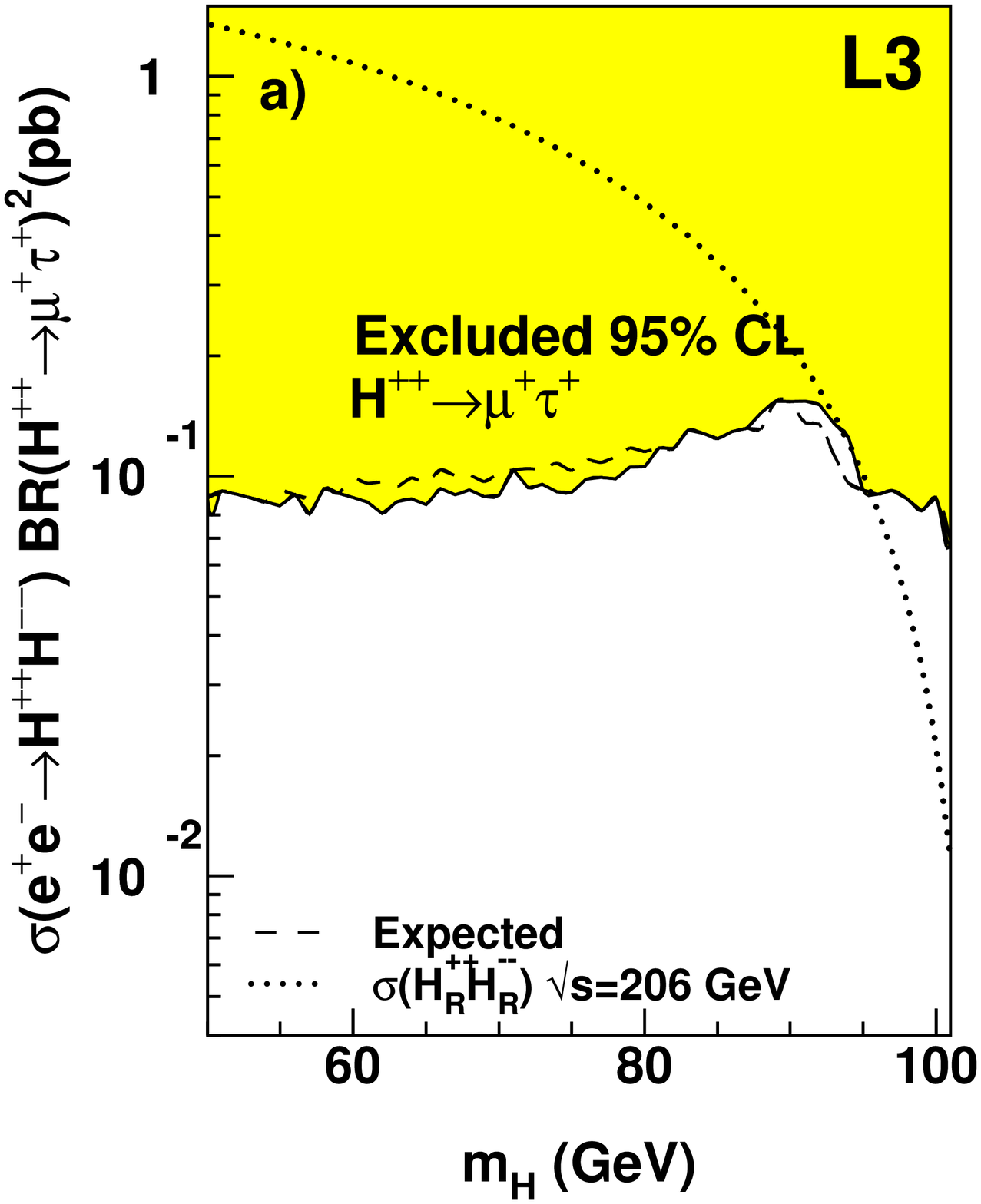,width=0.5\textwidth}}&
\mbox{\epsfig{figure=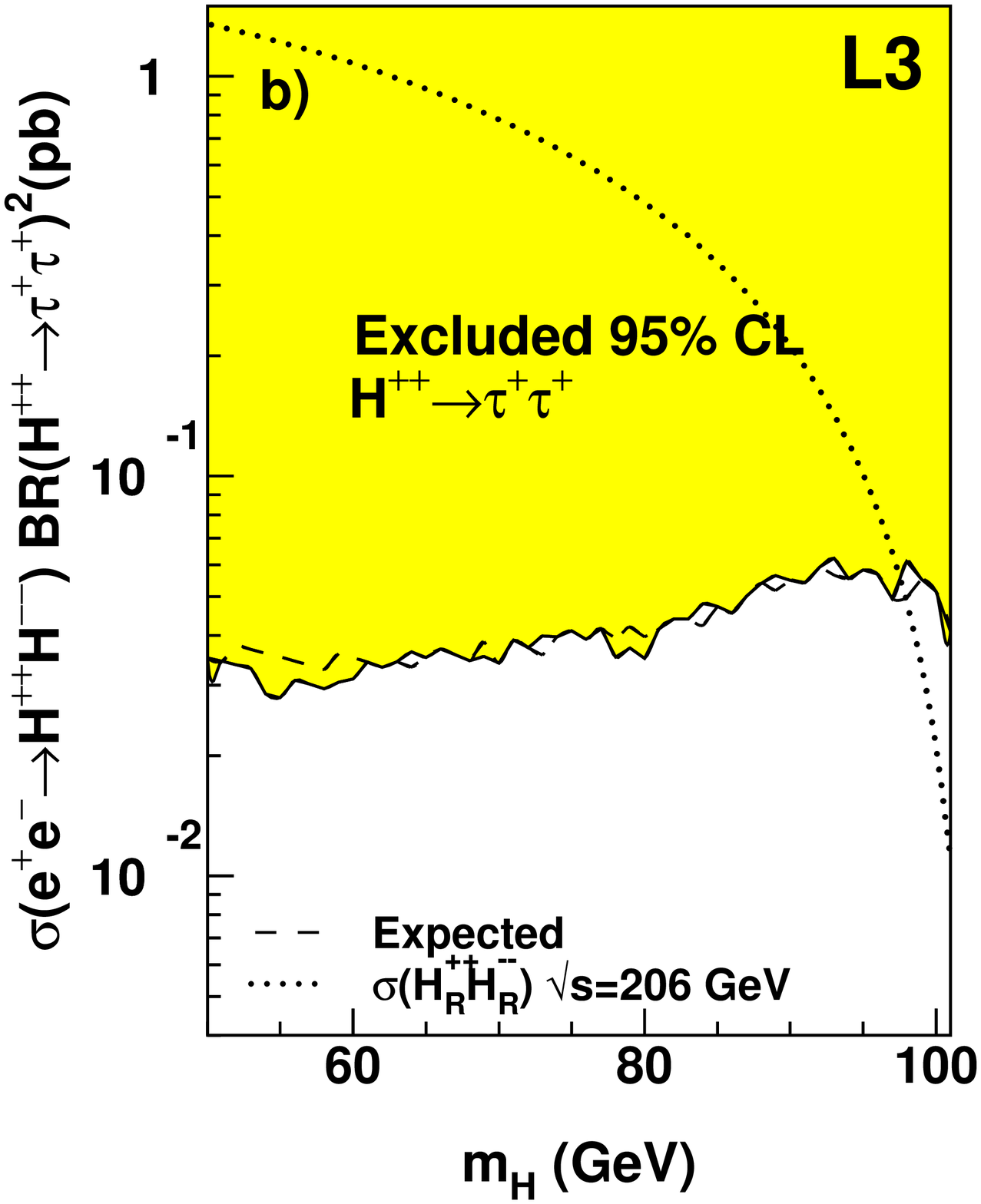,width=0.5\textwidth}}\\
\end{tabular}
\end{center}
\caption[]{\label{fig:5} Observed and expected limits on the cross
  section of doubly
  charged Higgs boson pair-production times its branching ratio in a given
  final state  as a function of $m_{\rm H}$ for the: a) $h_{\mu\tau}$ and
  b) $h_{\tau\tau}$
  couplings. The expected cross section for the $s$-channel
  production of a right-handed doubly-charged Higgs boson is also
  shown.}
\end{figure}

In the absence of a signal, upper limits on the production cross
section of doubly-charged Higgs bosons are derived as a function of
$m_{\rm H}$ and converted to lower limits on $m_{\rm H}$.  The
log-likelihood ratio technique~\cite{lnq} is used to calculate the
observed and expected 95\% confidence level cross section limits,
presented, as a function of $m_{\rm H}$ for the different couplings,
in Figures~\ref{fig:4} and~\ref{fig:5}. Cross sections between 0.1\,pb
and 0.01\,pb are excluded, depending on $m_{\rm H}$ and on the
coupling.

The limits include systematic uncertainties on the signal efficiency and
the background normalisation. These follow from uncertainties in the
determination of the energy scale of the detector, on the event
selection and lepton identification criteria, on Monte Carlo
statistics and on the cross section of the Standard Model background
processes. Table~\ref{table:syst} gives the total systematic
uncertainties for the different couplings. These uncertainties reduce
the sensitivity by a few hundred Me\kern -0.1em V.

Lower limits on $m_{\rm H}$ are extracted by comparing the cross
section upper limits with the known cross section of the process
$\epem\ra\rm H^{++}H^{--}$~\cite{Swartz:1989qz,Huitu:1998zt}. The most
conservative scenario of a right-handed $\rm H^{\pm\pm}$ and the absence of a
$t$-channel contribution to $\rm H^{\pm\pm}$ production is
considered. The observed limits vary from $95.5 \GeV$ to $100.2 \GeV$,
depending on the coupling and are listed in Table~\ref{table:limits}
together with the expected ones.

\section*{Constraints from Bhabha Scattering}

The measurements of the cross sections and forward-backward asymmetries
of the $\rm \epem \ra \epem$ process in $\rm 243.7\,pb^{-1}$
of data at $\sqrt{s}=130-189\GeV$ are described in
Reference~\citen{fpp} and found to be in good agreement with the
Standard Model predictions~\cite{ZFITTER,TOPAZ0}.  Similar analyses
are applied to $\rm 446.8\,pb^{-1}$ of data collected at
$\sqrt{s}=192-209\GeV$. The results are also in good agreement with
the Standard Model predictions, and show no evidence for the exchange
of a doubly-charged Higgs boson.

\begin{figure}[hp]
\begin{center}
\mbox{\epsfig{figure=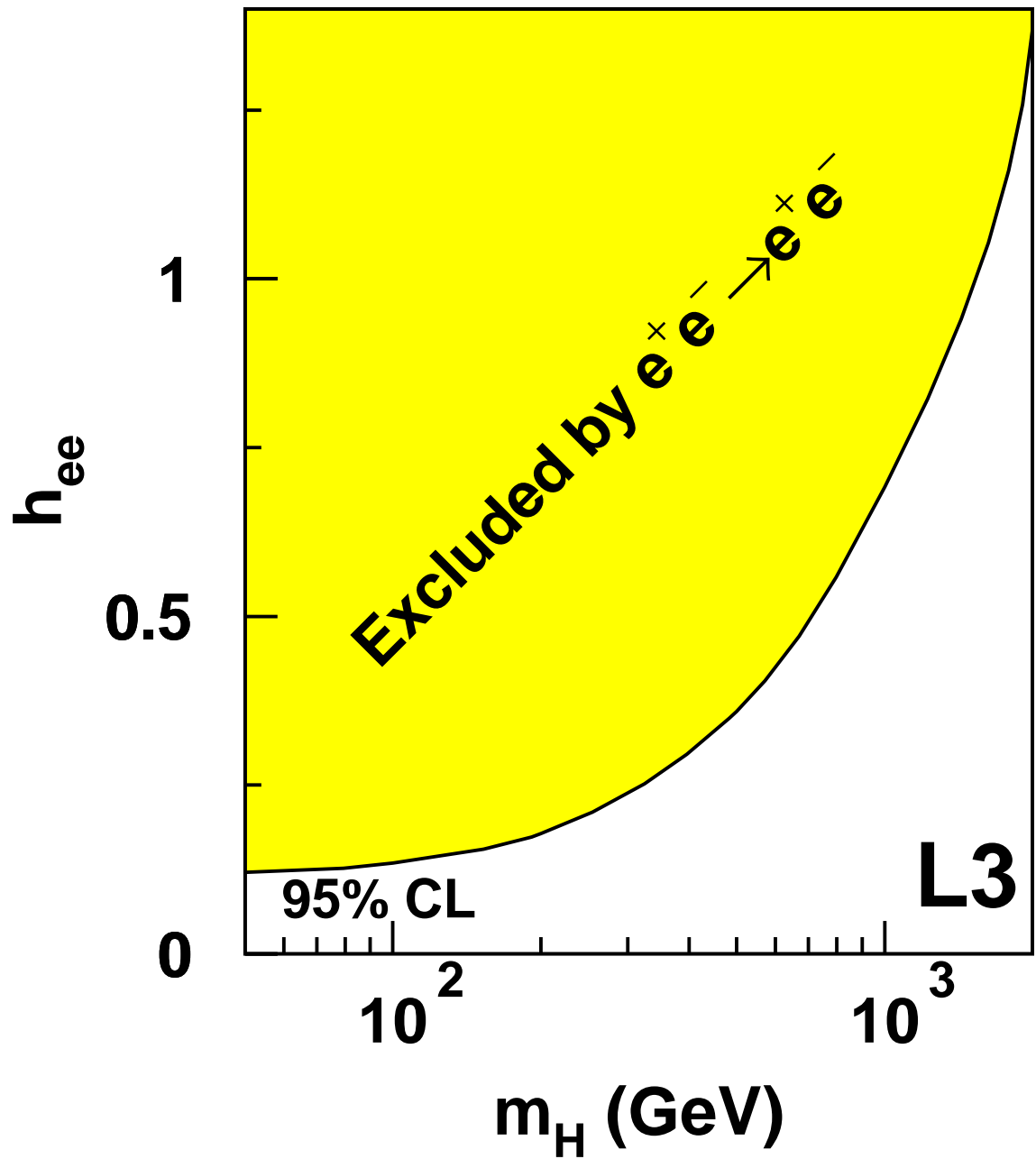,height=0.8\textheight}}
\end{center}
\caption[]{\label{fig:6} 
Region in the $h_{\rm ee}$ {\it vs.} $m_{\rm H}$  plane excluded
by the study of the $\epem \ra \epem$ process.}
\end{figure}

A fit for $h_{\rm ee}$ is performed to the measured cross
sections and forward-backward asymmetries for $\sqrt{s}=130-209\GeV$
and several hypotheses on the value of $m_{\rm H}$.  Experimental
systematic uncertainties~\cite{fpp} and uncertainties on the Standard
Model predictions~\cite{kobel} are taken into account in the
fit. Upper limits on $h_{\rm ee}$ at 95\% confidence level are derived
as a function of $m_{\rm H}$ and shown in Figure~\ref{fig:6}. The
exclusion region for $h_{\rm ee}>0.7$ extends to the Te\kern -0.1em V
scale and is complementary to that derived here from the search for
pair-production of doubly-charged Higgs bosons.


\bibliographystyle{l3stylem}

\newpage

%
\newpage
\section*{Author List}
%
%
%
%
%
%

\newcount\tutecount  \tutecount=0
\def\tutenum#1{\global\advance\tutecount by 1 \xdef#1{\the\tutecount}}
\def\tute#1{$^{#1}$}
\tutenum\aachen            
\tutenum\nikhef            
\tutenum\mich              
\tutenum\lapp              
\tutenum\basel             
\tutenum\lsu               
\tutenum\beijing           
\tutenum\bologna           
\tutenum\tata              
\tutenum\ne                
\tutenum\bucharest         
\tutenum\budapest          
\tutenum\mit               
\tutenum\panjab            
\tutenum\debrecen          
\tutenum\dublin            
\tutenum\florence          
\tutenum\cern              
\tutenum\wl                
\tutenum\geneva            
\tutenum\hefei             
\tutenum\lausanne          
\tutenum\lyon              
\tutenum\madrid            
\tutenum\florida           
\tutenum\milan             
\tutenum\moscow            
\tutenum\naples            
\tutenum\cyprus            
\tutenum\nymegen           
\tutenum\caltech           
\tutenum\perugia           
\tutenum\peters            
\tutenum\cmu               
\tutenum\potenza           
\tutenum\prince            
\tutenum\riverside         
\tutenum\rome              
\tutenum\salerno           
\tutenum\ucsd              
\tutenum\sofia             
\tutenum\korea             
\tutenum\purdue            
\tutenum\psinst            
\tutenum\zeuthen           
\tutenum\eth               
\tutenum\hamburg           
\tutenum\taiwan            
\tutenum\tsinghua          

{
\parskip=0pt
\noindent
{\bf The L3 Collaboration:}
\ifx\selectfont\undefined
 \baselineskip=10.8pt
 \baselineskip\baselinestretch\baselineskip
 \normalbaselineskip\baselineskip
 \ixpt
\else
 \fontsize{9}{10.8pt}\selectfont
\fi
\medskip
\tolerance=10000
\hbadness=5000
\raggedright
\hsize=162truemm\hoffset=0mm
\def\r{\rlap,}
\noindent

P.Achard\r\tute\geneva\ 
O.Adriani\r\tute{\florence}\ 
M.Aguilar-Benitez\r\tute\madrid\ 
J.Alcaraz\r\tute{\madrid}\ 
G.Alemanni\r\tute\lausanne\
J.Allaby\r\tute\cern\
A.Aloisio\r\tute\naples\ 
M.G.Alviggi\r\tute\naples\
H.Anderhub\r\tute\eth\ 
V.P.Andreev\r\tute{\lsu,\peters}\
% ...
\newpage
\begin{list}{A}{\itemsep=0pt plus 0pt minus 0pt\parsep=0pt plus 0pt minus 0pt
                \topsep=0pt plus 0pt minus 0pt}
\item[\aachen]
 III. Physikalisches Institut, RWTH, D-52056 Aachen, Germany$^{\S}$
\item[\nikhef] National Institute for High Energy Physics, NIKHEF, 
     and University of Amsterdam, NL-1009 DB Amsterdam, The Netherlands
\item[\mich] University of Michigan, Ann Arbor, MI 48109, USA
\item[\lapp] Laboratoire d'Annecy-le-Vieux de Physique des Particules, 
     LAPP,IN2P3-CNRS, BP 110, F-74941 Annecy-le-Vieux CEDEX, France
\item[\basel] Institute of Physics, University of Basel, CH-4056 Basel,
     Switzerland
\item[\lsu] Louisiana State University, Baton Rouge, LA 70803, USA
\item[\beijing] Institute of High Energy Physics, IHEP, 
  100039 Beijing, China$^{\triangle}$ 
\item[\bologna] University of Bologna and INFN-Sezione di Bologna, 
     I-40126 Bologna, Italy
\item[\tata] Tata Institute of Fundamental Research, Mumbai (Bombay) 400 005, India
\item[\ne] Northeastern University, Boston, MA 02115, USA
\item[\bucharest] Institute of Atomic Physics and University of Bucharest,
     R-76900 Bucharest, Romania
\item[\budapest] Central Research Institute for Physics of the 
     Hungarian Academy of Sciences, H-1525 Budapest 114, Hungary$^{\ddag}$
\item[\mit] Massachusetts Institute of Technology, Cambridge, MA 02139, USA
\item[\panjab] Panjab University, Chandigarh 160 014, India.
\item[\debrecen] KLTE-ATOMKI, H-4010 Debrecen, Hungary$^\P$
\item[\dublin] Department of Experimental Physics,
  University College Dublin, Belfield, Dublin 4, Ireland
\item[\florence] INFN Sezione di Firenze and University of Florence, 
     I-50125 Florence, Italy
\item[\cern] European Laboratory for Particle Physics, CERN, 
     CH-1211 Geneva 23, Switzerland
\item[\wl] World Laboratory, FBLJA  Project, CH-1211 Geneva 23, Switzerland
\item[\geneva] University of Geneva, CH-1211 Geneva 4, Switzerland
\item[\hefei] Chinese University of Science and Technology, USTC,
      Hefei, Anhui 230 029, China$^{\triangle}$
\item[\lausanne] University of Lausanne, CH-1015 Lausanne, Switzerland
\item[\lyon] Institut de Physique Nucl\'eaire de Lyon, 
     IN2P3-CNRS,Universit\'e Claude Bernard, 
     F-69622 Villeurbanne, France
\item[\madrid] Centro de Investigaciones Energ{\'e}ticas, 
     Medioambientales y Tecnol\'ogicas, CIEMAT, E-28040 Madrid,
     Spain${\flat}$ 
\item[\florida] Florida Institute of Technology, Melbourne, FL 32901, USA
\item[\milan] INFN-Sezione di Milano, I-20133 Milan, Italy
\item[\moscow] Institute of Theoretical and Experimental Physics, ITEP, 
     Moscow, Russia
\item[\naples] INFN-Sezione di Napoli and University of Naples, 
     I-80125 Naples, Italy
\item[\cyprus] Department of Physics, University of Cyprus,
     Nicosia, Cyprus
\item[\nymegen] University of Nijmegen and NIKHEF, 
     NL-6525 ED Nijmegen, The Netherlands
\item[\caltech] California Institute of Technology, Pasadena, CA 91125, USA
\item[\perugia] INFN-Sezione di Perugia and Universit\`a Degli 
     Studi di Perugia, I-06100 Perugia, Italy   
\item[\peters] Nuclear Physics Institute, St. Petersburg, Russia
\item[\cmu] Carnegie Mellon University, Pittsburgh, PA 15213, USA
\item[\potenza] INFN-Sezione di Napoli and University of Potenza, 
     I-85100 Potenza, Italy
\item[\prince] Princeton University, Princeton, NJ 08544, USA
\item[\riverside] University of Californa, Riverside, CA 92521, USA
\item[\rome] INFN-Sezione di Roma and University of Rome, ``La Sapienza",
     I-00185 Rome, Italy
\item[\salerno] University and INFN, Salerno, I-84100 Salerno, Italy
\item[\ucsd] University of California, San Diego, CA 92093, USA
\item[\sofia] Bulgarian Academy of Sciences, Central Lab.~of 
     Mechatronics and Instrumentation, BU-1113 Sofia, Bulgaria
\item[\korea]  The Center for High Energy Physics, 
     Kyungpook National University, 702-701 Taegu, Republic of Korea
\item[\purdue] Purdue University, West Lafayette, IN 47907, USA
\item[\psinst] Paul Scherrer Institut, PSI, CH-5232 Villigen, Switzerland
\item[\zeuthen] DESY, D-15738 Zeuthen, Germany
\item[\eth] Eidgen\"ossische Technische Hochschule, ETH Z\"urich,
     CH-8093 Z\"urich, Switzerland
\item[\hamburg] University of Hamburg, D-22761 Hamburg, Germany
\item[\taiwan] National Central University, Chung-Li, Taiwan, China
\item[\tsinghua] Department of Physics, National Tsing Hua University,
      Taiwan, China
\item[\S]  Supported by the German Bundesministerium 
        f\"ur Bildung, Wissenschaft, Forschung und Technologie
\item[\ddag] Supported by the Hungarian OTKA fund under contract
numbers T019181, F023259 and T037350.
\item[\P] Also supported by the Hungarian OTKA fund under contract
  number T026178.
\item[$\flat$] Supported also by the Comisi\'on Interministerial de Ciencia y 
        Tecnolog{\'\i}a.
\item[$\sharp$] Also supported by CONICET and Universidad Nacional de La Plata,
        CC 67, 1900 La Plata, Argentina.
\item[$\triangle$] Supported by the National Natural Science
  Foundation of China.
\end{list}
}
\vfill



\begin{table}
\begin{center}
\begin{tabular}{|l|ccccccc|}
\hline
$\sqrt{s}$ (\GeV)      & 188.6 & 191.6 & 195.5 & 199.5 & 201.7 & 205.0 & 206.6 \\
Luminosity (pb$^{-1}$) & 176.8 & \pho29.8 &  \pho84.1 &  \pho84.0 &  \pho39.2 &  \pho80.0 & 130.2 \\
\hline
\end{tabular}
\caption{Average centre-of-mass energies and corresponding integrated luminosities.}
\label{table:lumi}
\end{center}
\end{table}

\begin{table}
\begin{center}
\begin{tabular}{|c|c|l|}
\hline
Coupling & $\rm H^{++}H^{--}\ra$  & Analyses \\
\hline
$h_{\rm ee}       $& $\rm e^+e^+e^-e^-            $& e\,e\,e\,e, e\,e\,e\,$\gamma$\\
$h_{\rm e\mu}     $& $\rm e^+\mu^+e^-\mu^-        $& e\,e$\,\mu\,\mu$, e\,$\gamma\,\mu\,\mu$ \\
$h_{\rm e\tau}    $& $\rm e^+\tau^+e^-\tau^-      $& e\,e$\,\tau\,\tau$, e\,e\,jet\,jet, e\,$\gamma$\,jet\,jet\\
$h_{\rm \mu\mu}   $& $\rm \mu^+\mu^+\mu^-\mu^-    $& $\mu\,\mu\,\mu\,\mu$, $\mu\,\mu\,\mu$\,MIP\\
$h_{\rm \mu\tau}  $& $\rm \mu^+\tau^+\mu^-\tau^-  $& $\mu\,\mu$\,jet\,jet, $\mu\,\mu\,\tau\,\tau$\\
$h_{\rm \tau\tau} $& $\rm \tau^+\tau^+\tau^-\tau^-$& e\,e$\,\tau\,\tau$, e\,e\,jet\,jet, e\,$\gamma$\,jet\,jet, $\mu\,\mu$\,jet\,jet, $\mu\,\mu\,\tau\,\tau$, $\tau\,\tau\,\tau\,\tau$\\
\hline
\end{tabular}
\caption{Analyses used for the different couplings and the corresponding
  final states.}
\label{table:analyses}
\end{center}
\end{table}

\begin{table}
\begin{center}
\begin{tabular}{|c|c|c|c|c|c|c|c|}
\hline
 & \multicolumn{3}{c|}{Preselection} & \multicolumn{4}{|c|}{Final results}\\
\cline{2-8}
Coupling & $N_D$ & $N_B$ & $N_S$ & $N_D$ & $N_B$ & $N_S$ &
 $\varepsilon$ (\%) \\
\hline
$h_{\rm ee}       $& \phantom{000}7 & \phantom{00}10.9 &           18.3 & \phantom{0}0 & \phantom{0}2.7 &           16.9 & $46 - 63$\\
$h_{\rm e\mu}     $& \phantom{00}12 & \phantom{00}10.2 &           12.9 & \phantom{0}9 & \phantom{0}6.5 &           12.4 & $35 - 44$\\
$h_{\rm e\tau}    $&           1308 &           1250\phantom{.0} & \phantom{0}7.5 &           23 &           21.9 & \phantom{0}6.5 & $39 - 44$\\
$h_{\rm \mu\mu}   $& \phantom{000}0 & \phantom{000}1.0 &           10.6 & \phantom{0}0 & \phantom{0}0.7 & \phantom{0}9.2 & $28 - 32$\\
$h_{\rm \mu\tau}  $& \phantom{000}8 & \phantom{000}4.4 & \phantom{0}8.2 & \phantom{0}3 & \phantom{0}4.3 & \phantom{0}4.7 & $19 - 22$\\
$h_{\rm \tau\tau} $&           1318 &           1258\phantom{.0} &           12.5 &           28 &           27.1 &           11.1 & $46 - 53$\\
\hline
\end{tabular}
\caption{Numbers of events observed in data, $N_D$,  expected from
  Standard Model processes, $N_B$, and for a $m_{\rm H}=95 \GeV$
  signal, $N_S$, after the application of the preselection
  and  final selection cuts. Final selection efficiencies,
  $\varepsilon$, for $m_{\rm H}=60-100\GeV$ are also given.}
\label{table:selection}
\end{center}
\end{table}

\begin{table}
\begin{center}
\begin{tabular}{|c|c|c|}
\hline
Coupling & Signal (\%) & Background (\%)\\
\hline
$h_{\rm ee}       $& 1.8 &           16.8\\
$h_{\rm e\mu}     $& 1.8 &           14.5\\
$h_{\rm e\tau}    $& 1.8 & \phantom{0}9.3\\
$h_{\rm \mu\mu}   $& 1.8 &           15.1\\
$h_{\rm \mu\tau}  $& 1.4 &           10.7\\
$h_{\rm \tau\tau} $& 3.2 &           10.4\\
\hline
\end{tabular}
\caption{Systematic uncertainties on the signal efficiencies and on the
  background levels for the different couplings.}
\label{table:syst}
\end{center}
\end{table}

\begin{table}
\begin{center}
\begin{tabular}{|c|c|c|}
\hline
Coupling & Observed (Ge\kern -0.1em V)&  Expected (Ge\kern -0.1em V)\\
\hline
$h_{\rm ee}       $&           100.2 &           100.1 \\
$h_{\rm e\mu}     $& \phantom{0}99.8 & \phantom{0}99.7 \\
$h_{\rm e\tau}    $& \phantom{0}97.2 & \phantom{0}95.5 \\
$h_{\rm \mu\mu}   $& \phantom{0}99.4 & \phantom{0}99.1 \\
$h_{\rm \mu\tau}  $& \phantom{0}95.5 & \phantom{0}93.8 \\
$h_{\rm \tau\tau} $& \phantom{0}97.3 & \phantom{0}97.6 \\
\hline
\end{tabular}
\caption{Observed and expected limits on $m_{\rm H}$ at 95\%
  confidence level.}
\label{table:limits}
\end{center}
\end{table}


\newpage

\end{document}